\documentclass[%
 reprint,
 amsmath,amssymb,
 aps,
]{revtex4-2}

\usepackage{graphicx}
\usepackage{dcolumn}
\usepackage{bm}
\usepackage{amsmath}

\usepackage{diagbox}
\usepackage{pifont}
\usepackage{multirow}
\usepackage{ctable}
\usepackage{float}

\begin{document}

\preprint{APS/123-QED}








\title{Granger Causality for Compressively Sensed Sparse Signals}

\author{Aditi Kathpalia}
\email{kathpalia@cs.cas.cz}
 \affiliation{%
 Department of Complex Systems, \\
 Institute of Computer Science of the Czech Academy of Sciences, \\
 Prague, Czech Republic
}
 \author{Nithin Nagaraj}%
\email{nithin@nias.res.in}
\affiliation{%
 Consciousness Studies Programme, National Institute of Advanced Studies, \\ 
 Bengaluru, India
}%




\date{\today}

\begin{abstract}
Compressed sensing is a scheme that allows for sparse signals to be acquired, transmitted and stored using far fewer measurements than done by conventional means employing Nyquist sampling theorem. Since many naturally occurring signals are sparse (in some domain), compressed sensing has rapidly seen popularity in a number of applied physics and engineering applications, particularly in designing signal and image acquisition strategies, e.g., magnetic resonance imaging, quantum state tomography, scanning tunneling microscopy, analog to digital conversion technologies. Contemporaneously, causal inference has become an important tool for the analysis and understanding of processes and their interactions in many disciplines of science, especially those dealing with complex systems. Direct causal analysis for compressively sensed data is required to avoid the task of reconstructing the compressed data. Also, for some sparse signals, such as for sparse temporal data, it may be difficult to discover causal relations directly using available data-driven/ model-free causality estimation techniques. In this work, we provide a mathematical proof that structured compressed sensing matrices, specifically Circulant and Toeplitz, preserve causal relationships in the compressed signal domain, as measured by Granger Causality. We then verify this theorem on a number of bivariate and multivariate coupled sparse signal simulations which are compressed using these matrices. We also demonstrate a real world application of network causal connectivity estimation from sparse neural spike train recordings from rat prefrontal cortex.
\end{abstract}

\maketitle


\section{\label{sec:level1}Introduction}

The use of compressed (or compressive) sensing or sampling for acquisition of signals has become very popular in the last two decades. The basic idea behind the technique is the acquisition and compression of data at the same time. Since most real-world signals are sparse in some domain, they can be acquired with only a few measurements. These measurements contain sufficient information about the signals, making their perfect reconstruction possible (assuming sparsity)~\cite{donoho2006compressed, candes2008introduction}.

Compressed sensing has found applications in designing magnetic resonance imaging techniques~\cite{lustig2007sparse, lustig2008compressed}, cameras~\cite{takhar2006compressed}, analog to digital conversion technologies~\cite{candes2008introduction}, ghost imaging~\cite{katz2009compressive, zhao2012ghost}, nuclear magnetic resonance spectroscopy~\cite{kazimierczuk2011accelerated, holland2011fast}, quantum state tomography~\cite{gross2010quantum, flammia2012quantum}, radio interferometry based black hole observation~\cite{honma2014super, ikeda2016precl} and scanning tunneling microscopy~\cite{nakanishi2016compressed}. Since large number of signals are being acquired and stored in the compressed domain, it becomes imperative to analyze the properties of these signals and study their interdependence in the compressed domain itself, avoiding the cumbersome task of their reconstruction. Hence, methods have been proposed to perform many important computations in the domain of signal processing and learning on compressed low dimensional data directly. These computations include tasks such as regression~\cite{zhou2009compressed}, classification~\cite{blum2005random, haupt2006compressive, davenport2007smashed, duarte2007multiscale}, signal detection~\cite{duarte2006sparse}, nearest neighbor finding~\cite{indyk1998approximate} and manifold learning~\cite{hegde2007random}.

Causality analysis techniques study interactions between different variables of a complex system and provides richer information than mere linear correlation based analysis.
This information is required by many disciplines of science in order to study specific systems based on cause-effect relationships between observables and eventually control their required parts, where need be~\cite{pearl2018book, kathpalia2021measuring}. Neuroscience~\cite{seth2015granger, vicente2011transfer}, earth sciences~\cite{mosedale2006granger, tirabassi2015study, runge2019inferring, paluvs2014multiscale, jajcay2018synchronization}, econometrics~\cite{geweke1984inference, hiemstra1994testing, chiou2008economic, dimpfl2013using}, engineering~\cite{bauer2007finding} etc., have employed different time series causality estimation techniques to study interactions in systems such as the human brain, stock prices, earth's climate. 

Causality estimation for time series data began with the pioneering work of Granger~\cite{granger1969investigating}. Granger Causality (GC) continues  to be the most widely used method for causality estimation till date. GC is inspired from Wiener's idea of causal prediction, according to which ``a time series $X$ causes a time series $Z$, if the past values of $X$ contain information that help predict $Z$ above and beyond the information contained in the past values of $Z$ alone''. To estimate GC from $X$ to $Z$, $Z$ is modelled as a linear autoregressive process in two independent ways: 
\begin{equation}
    Z(t)=\sum_{\tau=1}^{\infty}(a_\tau Z(t-\tau))+\sum_{\tau=1}^{\infty}(c_\tau X(t-\tau))+\varepsilon_c,
    \label{eq_granger_causal_model}
\end{equation}
    
\begin{equation}
Z(t)=\sum_{\tau=1}^{\infty}(b_\tau Z(t-\tau))+\varepsilon,
 \label{eq_granger_noncausal_model}
\end{equation}
where $t$ denotes any time instance,  $a_\tau, b_\tau, c_\tau$ are coefficients at a time lag of $\tau$ and  $\varepsilon_c, \varepsilon$ are error terms (gaussian distributed) in the two models. To determine if $X$ causes $Z$ or not, logarithm of the ratio of the prediction error variances is computed:
\begin{equation}
    F_{X \rightarrow Y}=\ln \frac{var(\varepsilon)}{var(\varepsilon_c)}.
\end{equation}
This measure is called the GC F-statistic. If the model represented by Eq.~(\ref{eq_granger_causal_model}) is a better model for $Z(t)$ than Eq.~(\ref{eq_granger_noncausal_model}), then var($\varepsilon_c$) $<$ var($\varepsilon$) and $F_{X \rightarrow Y}$ will be greater than 0, suggesting that $X$ \emph{Granger Causes} $Z$. In spite of using a simplistic autoregressive model, GC is applicable (in principle) in a wide range of covariance stationary processes~\cite{mvgc, anderson2011statistical}.

A number of attempts have been made to generalize GC to the nonlinear case, such as, by using, an estimator based on correlation integral~\cite{hiemstra1994testing}, a non-parametric regression approach~\cite{bell1996non}, local linear predictors~\cite{chen2004analyzing}, mutual nearest neighbors~\cite{schiff1996detecting, le1999nonlinear}, kernel estimators~\cite{marinazzo}, just to name a few. Other causality estimation techniques have also been proposed based on the Wiener's idea of causation. These include Transfer Entropy~\cite{schreiber}, Conditional Mutual Information~\cite{paluvs2001synchronization}, Compression-Complexity Causality~\cite{kathpalia2019data}, and Direct Transfer Function~\cite{kaminski2001evaluating}. All these techniques are largely model-free/non-parametric and do not require an explicit assumption of the underlying causal-model. Other than these, there are other techniques which are model-based, which require an explicit assumption of the underlying physical model generating the data. Owing to these assumptions, these methods are often specific to certain systems in specific application domains. Some examples include Dynamic Causal Modeling~\cite{friston} which was developed for neuroscience and Structural Equation Modeling~\cite{pearl2000causality} which has found applications in medicine, engineering and social sciences.

In today's world of data-driven science, both compressed sensing and causality estimation have gained impressive ground across disciplines, and play an important role in experimental~\cite{chan2008single, gross2010quantum, studer2012compressive, tayler2014ultrafast, ma2021high} and computational~\cite{andrade2012application, saigre2022intelligent, tsumori2020contour} sciences as well as the study of complex systems and networks~\cite{barranca2016compressive, sheikhattar2018extracting, paluvs2007nonlinearity, seth2015granger, finkle2018windowed, runge2019inferring}. By saving measurement time and cost, compressed sensing has enabled experiments to be conducted more efficiently. Causality estimation, on the other hand, has supplied tools to study interactions in systems and unravel their underlying dynamics. The motivation for this study lies at the intersection of these two disciplines.
Specifically, we wish to extend causal estimation to the compressed domain. It is often the case that causal network discovery is required to be done on high dimensional data having large number of variables. Application of causality estimation techniques to compressed signals will prevent the computationally expensive task of reconstructing signals acquired using compressive sensing. This would enable a fast and reliable causal discovery, helping in \emph{compressed computation}, which has gained importance in recent times. In the following subsections, we discuss the technique of compressed sensing and the requirement for causal inference in the compressive domain in some detail.

\subsection{Compressed Sensing}

Compressed sensing is a relatively recent signal processing technique that helps in the acquisition, transmission, storage as well as modification of large amounts of data in an efficient manner. It is well-known that when a signal is sampled at the Nyquist rate, perfect recovery of the original signal is guaranteed. However, the theory of compressed sensing claims that signal recovery is possible by using far fewer measurements under certain conditions. There are two conditions required for this: (1) the input signal be \emph{sparse} when expressed in a proper basis, and (2) the basis in which it is sensed/acquired be \emph{incoherent} with the above mentioned sparsifying basis~\cite{candes2008introduction, candes2006robust}. \emph{Sparsity} refers to the case in which most entries of a discrete-time signal/ array are zero. If a signal has `$k$' non-zero entries, then $k$ is much less than the total length of the signal. This means that, when expressed in the proper basis $\psi$, which is called the sparsifying matrix, these signals have concise representations. \emph{Incoherence} is basically the idea that objects that have a sparse representation in $\psi$ should spread out in the domain in
which they are acquired (or sensed). This is analogous to the property of {\it duality} -- an impulse in the time domain is spread out in the frequency domain (and vice versa). In other words, incoherence means that the sampling/sensing waveforms have a very dense representation in $\psi$.

Let $y$ (an $m \times 1$ dimension vector) be a set of measurements acquired for a sparse signal $z$ (of dimension $n \times 1$). $z$ could be itself sparse or could be expressed in the form $\psi \alpha$, where $\psi$ is the sparsifying matrix and $\alpha$ contains the sparse coefficients. $y$ is called the compressed signal or compressive measurement vector. Then, sparse signal recovery can be seen as a compressed sensing problem as follows:
\begin{equation}
    y = \phi z,
    \label{eq_CS_1}
\end{equation}
where $\phi$ is an $m \times n$ sensing matrix. The input vector $z$ is in $\mathbb{R}^n$ and is mapped to the output vector $y$, which is in $\mathbb{R}^m$. The sensing matrix, $\phi$, has a rank $\leq m$ (and $m$ $\ll$ $n$).

$\phi$ must satisfy, $\phi z_1 \neq \phi z_2$ for all $k$-sparse vectors $z_1 \neq z_2$. Here, $k$-sparse vectors are those which contain a maximum number of $k$ non-zero entries. This property allows the mapping to be invertible on all $k-$sparse vectors, ensuring that no two output vectors are mapped to the exact same sparse vector during recovery. For this to be the case, it is required that $\phi$ has at least $2k$ rows, i.e. $m \geq 2k$. 

\emph{Restricted Isometry Property} (RIP)~\cite{candes2008restricted} is a key notion in compressed sensing that determines the efficiency and robustness of the sensing matrix $\phi$ to capture information about sparse signals, such that the aim of recovering them is satisfactorily fulfilled. For each integer, $k=1,2,\ldots$, let $\delta_k$ be the RIP/isometry constant of matrix $\phi$ associated with that $k$. Then, $\delta_k$ is defined as the smallest positive value such that
\begin{equation}
    (1-\delta_k)\left\Vert z \right\Vert_{l_2}^{2} \leq \left\Vert \phi z \right\Vert_{l_2}^{2} \leq (1+\delta_k)\left\Vert z \right\Vert_{l_2}^{2},
    \label{eq_CS_RIP}
\end{equation}
holds for all $k$-sparse vectors $z$\footnote{$\left\Vert u \right\Vert_{l_p}$ denotes the $l_p$ norm of vector $u$, which is computed as $\left\Vert u \right\Vert_{l_p}=(\sum_{i=1}^{n}{|u_i|}^p)^{1/p}$, where $n$ is the length of the vector.}. $\phi$ is said to fulfill an RIP of order $k$, if $\delta_k$ is not too close to one. This implies that for $k$-sparse signals, the euclidean length is roughly preserved, and so these $k$-sparse vectors cannot be in the null space of $\phi$. Consequently, this ensures that their reconstruction is possible. 

Even when $\phi$ satisfies the above conditions, Eq.~\ref{eq_CS_1} still remains under-determined with infinitely many solutions. The best (sparse) solution can be recovered from the compressed measurement vector $y$, using a variety of algorithms, which employ, for example, optimization techniques (such as $l_1$ minimization~\cite{donoho2006compressed, tsaig2006extensions}) or greedy approach (such as Orthogonal Matching Pursuit~\cite{tropp2007signal}). 

\subsection{Causal Inference in the Compressive Domain}

As compressed sensing schemes are being widely employed for the acquisition of signals, their transmission and storage, it is important to know whether causal relationships between different variables in the compressed domain can be inferred directly from the compressed data, that is without reconstructing the signals. Also, it is necessary to know, under what circumstances and for what kind of sensing matrices are causal relationships preserved in the compressed domain. 

Further, sometimes, in case of sparse data, available techniques for model-free causality estimation may not perform adequately when applied directly to the original data (uncompressed). For example, Granger Causality cannot be computed directly for point processes and only spectral GC estimated from fourier transformed stationary point processes is meaningful~\cite{nedungadi2009analyzing}. Another way that has been explored for such cases is converting the point process data to continuous domain through the use of smoothing kernels~\cite{sameshima1999using} or low pass filters~\cite{kaminski2001evaluating, zhu2003probing} and then applying GC or one of its extensions. Other than this, GC in combination with machine learning algorithms has also been used to learn causality for certain kinds of point processes~\cite{xu2016learning}. Transfer entropy (in its original discrete time form), when applied to point processes or event-based data, has also been shown to have a number of drawbacks and hence continuous time Transfer Entropy has been developed and proposed as an alternative~\cite{spinney2017transfer, shorten2021estimating}. However, there are no works validating the use of model-free causal estimation techniques, generally on sparse processes or compressed representations of sparse processes. There are some specialized model-based methods that have been proposed to be applied to specific kinds of sparse processes. For example, point process generalized linear model (GLM) based techniques~\cite{truccolo2005point, okatan2005analyzing, kim2011granger, chen2010statistical, zhang2016statistical, casile2021robust}, MVAR nonlinear Poisson model~\cite{krumin2010multivariate}, dynamic Bayesian network~\cite{eldawlatly2010use} and Cox model~\cite{berry2012detecting} have been proposed for effective connectivity analysis of neural spike train data. 

In this work, we discuss, if and how the use of Granger causality can be done in order to discover causal relations for sparse data by using them directly as acquired in the compressive domain or transforming them into compressed domain. Specifically we show that structured compressed sensing matrices, circulant and toeplitz, when used to directly sense or (posteriorly) compress sparse signals preserve GC between these signals in the compressed domain. To the best of our knowledge, both our problem statement and the solution proposed are novel and no work has previously been done in this direction. 

When the connections in the network are sparse, there exist techniques such as Lasso Granger~\cite{arnold2007temporal}, its several variations~\cite{shojaie2010discovering, lozano2009grouped}, CaSPIAN~\cite{emad2014caspian} etc., that employ a combination of compressed sensing techniques (such as variable (model) selection, for example, Lasso) and causality estimation methods (such as GC). These are methods to infer network connectivity and are of use for study of systems such as Genetic Regulatory Networks~\cite{hlavavckova2015lasso}. These methods are not to be confused with the work that we present in this paper since none of these methods attempt to infer causal relationships in the compressed domain. 

This paper is organized as follows. In Section~\ref{section2}, we provide a mathematical proof that circulant and toeplitz structured sensing matrices preserve GC for compressively sensed signals. In Section~\ref{section_res}, we demonstrate the performance of these sensing matrices by discovering causality from signals sensed using these matrices. This is done for the case of bivariate and multivariate simulations of sparse signals. We also apply the matrices to estimate GC from compressed counterparts of neural spike trains recorded from rat prefrontal cortex. We discuss our results and conclude in Section~\ref{section_discussion}.

\section{Special Sensing Matrices that Preserve Granger Causality}
\label{section2}
As discussed before, a sensing matrix, $\phi$, on multiplication with an $n \times 1$ sparse signal, $z$, yields a compressed signal, $y$, of size $m \times 1$. This is according to the equation, $y=\phi z$. Gaussian random values in $\phi$ prove to be one of the most effective ways in helping recover $z$ from $y$~\cite{candes2006near, baraniuk2008simple}. However, one of the drawbacks of using a random matrix is that hardly any properties of the original signal are preserved after compression. Hence, it was intuitive to explore matrices which possess some kind of structure and also satisfy the essential properties for sparse signal recovery, so that no information is lost from the signals. 

Circulant and toeplitz matrices have been shown to be effective sensing matrices for recovery of sparse signals~\cite{rauhut2009circulant, yin2010practical}. In fact, toeplitz and partial random circulant matrices have both been shown to satisfy the RIP property~\cite{bajwa2007toeplitz, rauhut2012restricted}. In this section we provide complete mathematical proofs showing that signals sensed using either of these matrices are effective at preserving causality as measured by GC.

\subsection{Circulant Sensing Matrices}


Let $C$ be a circulant matrix of order $n \times n$, composed of random entries $a_0, a_1, a_2, \ldots, a_{n-1}$, where $a_i \sim \mathcal{N}(0,1)$ (Gaussian distribution with zero mean and unit variance):

\begin{equation}
C = 
\begin{bmatrix}
a_{0} & a_{n-1} & a_{n-2} & \cdots & a_{1} \\
a_{1} & a_{0} & a_{n-1} & \cdots & a_{2} \\
a_{2} & a_{1} & a_{0} & \cdots & a_{3} \\
\vdots  & \vdots & \vdots & \ddots & \vdots  \\
a_{n-1} & a_{n-2} & a_{n-3} & \cdots & a_{0}
\end{bmatrix}_{n \times n}.
\end{equation}

It is known that a circulant matrix is diagonalizable by a Discrete Fourier Transform (DFT) matrix~\cite{davis1979circulant}. Let $F_n$ denote the DFT matrix:
\begin{equation}
F_n = 
\begin{bmatrix}
1 & 1 & 1 & \cdots & 1 \\
1 & (\omega)^1 & (\omega)^2 & \cdots & (\omega)^{n-1} \\
1 & (\omega^2)^1 & (\omega^2)^2 & \cdots & (\omega^2)^{n-1} \\
\vdots  & \vdots & \vdots & \ddots & \vdots  \\
(w^{n-1})^1 & (w^{n-1})^2 & (w^{n-1})^3 & \cdots & (w^{n-1})^{n-1}
\end{bmatrix}_{n \times n}.
\end{equation}

So,
\begin{equation}
C_n =(F_{n}^{-1})DF_n,
\label{eq_circ_diag}
\end{equation}
where, $D$ is a diagonal matrix formed by the entries $D_{k,k}=G_k$, where $G_k$ is the $k^{th}$ coefficient of the DFT of $\vec{a}=[a_0, a_1, \ldots, a_{n-1}]$. Let, $z$ (of dimension $n \times 1$), when multiplied by $C$ return:


\begin{equation}
    y= C z
    \label{eq_CS_circ}
\end{equation}

From Eq.~(\ref{eq_circ_diag}),

\begin{equation}
\label{eq_circ_fourier1}
\begin{split}
    Cz &= ((F_n^{-1})DF_n)z, \\
    &= F_n^{-1}D(F_nz).
\end{split}
\end{equation}

So, if $Z(f_i)$ denotes the fourier coefficient of $z$ at frequency $f_i$,

\begin{equation}
\begin{split}
    Cz &= F_n^{-1}
\begin{bmatrix}
G_{0} & 0 & 0 & \cdots & 0 \\
0 & G_1 & 0 & \cdots & 0 \\
0 & 0 & G_2 & \cdots & 0 \\
\vdots  & \vdots & \vdots & \ddots & \vdots  \\
0 & 0 & 0 & \cdots & G_{n-1}
\end{bmatrix}
\begin{bmatrix}
Z(f_0) \\
Z(f_1) \\
Z(f_2) \\
\vdots \\
Z(f_{n-1})
\end{bmatrix},\\
    &= F_n^{-1}
\begin{bmatrix}
G_0Z(f_0) \\
G_1Z(f_1) \\
G_2Z(f_2) \\
\vdots \\
G_{n-1}Z(f_{n-1})
\end{bmatrix}.
\end{split}
\label{eq_full_circ_freqs}
\end{equation}

Multiplying eqn.~\ref{eq_CS_circ} with a projection matrix $P$ of order $m \times n$ (with $m<n$) on both sides.

\begin{equation}
    Py=PCz,
    \label{eq_full_circ_proj}
\end{equation}
where, 
\begin{equation}
P=
\begin{bmatrix}
1 & 0 & \cdots & 0 & 0 & \cdots & 0 \\
0 & 1 & \cdots & 0 & 0 & \cdots & 0 \\
\vdots  & \vdots & \ddots & \vdots & \vdots & \ddots & \vdots \\
0 & 0 & \cdots & 1 & 0 & \cdots & 0
\end{bmatrix}_{m \times n}.
\end{equation}
This is equivalent to multiplying $z$ with an $m \times n$ compressed sensing matrix $PC=C_{m \times n}$,  with circulant structure. From equations \ref{eq_full_circ_proj} and \ref{eq_full_circ_freqs},
\begin{equation}
\begin{split}
    Py &=PF_n^{-1}
    \begin{bmatrix}
    G_0Z(f_0) \\
    G_1Z(f_1) \\
    \vdots \\
    G_{n-1}Z(f_{n-1})
    \end{bmatrix} \\
    & = F_{m \times n}^{-1}
    \begin{bmatrix}
    G_0Z(f_0) \\
    G_1Z(f_1) \\
    \vdots \\
    G_{n-1}Z(f_{n-1})
    \end{bmatrix},
    \end{split}
    \label{eq_circ_proj_freqs}
\end{equation}
where,
\begin{equation}
    F_{m \times n}^{-1}=
    \begin{bmatrix}
    1 & 1 & 1 & \cdots & 1 \\
    1 & (\omega)^1 & (\omega^2)^1 & \cdots & (\omega^{n-1})^1 \\
    \vdots  & \vdots & \vdots & \ddots & \vdots  \\
    1 & (\omega)^{m-1} & (\omega^2)^{m-1} & \cdots & (\omega^{n-1})^{m-1}
    \end{bmatrix}.
\end{equation}
Let $Py =y'$, where $y'$ is an $m \times 1$ vector. $y'$ is essentially a compressively sensed vector. Multiplying \ref{eq_circ_proj_freqs} on both sides with Fourier transform matrix of order $m \times m$,

\begin{equation}
\begin{split}
    F_my' &=F_mF_{m \times n}^{-1}
    \begin{bmatrix}
    G_0Z(f_0) \\
    G_1Z(f_1) \\
    \vdots \\
    G_{n-1}Z(f_{n-1})
    \end{bmatrix}, \\
    &= P_{m \times n}
    \begin{bmatrix}
    G_0Z(f_0) \\
    G_1Z(f_1) \\
    \vdots \\
    G_{n-1}Z(f_{n-1})
    \end{bmatrix},
    \end{split}
\end{equation}
yields,
\begin{equation}
\label{eq_scaled_freq_comps}
\begin{split}
    \begin{bmatrix}
    Y'(f_0) \\
    Y'(f_1) \\
    \vdots \\
    Y'(f_{m-1})
    \end{bmatrix} &=
    \begin{bmatrix}
    G_0Z(f_0) \\
    G_1Z(f_1) \\
    \vdots \\
    G_{m-1}Z(f_{m-1})
    \end{bmatrix}, \\
    \begin{bmatrix}
    S_{y'y'}(f_0) \\
    S_{y'y'}(f_1) \\
    \vdots \\
    S_{y'y'}(f_{n-1})
    \end{bmatrix}
    &= 
    \begin{bmatrix}
    G_0^2S_{z,z}(f_0) \\
    G_1^2S_{z,z}(f_1) \\
    \vdots \\
    G_1^2S_{z,z}(f_{m-1})
    \end{bmatrix}.
    \end{split}
\end{equation}

Here, $Y'(f_0), Y'(f_1), \ldots$ are the fourier coefficients of $y'$. Squaring the equation on both sides, gives a relation between the spectral power coefficients of $y'$ and $z$. $S_{y',y'}$ and $S_{z,z}$ represent the spectral coefficients of $y'$ and $z$ respectively. 

For two processes $z_1$ and $z_2$, which may be coupled and can be modeled like autoregressive processes,

\begin{equation}
\begin{split}
    z_1(t) & =\sum_{j=1}^{\infty} {b_{11,j}z_1(t-j)}+\sum_{j=1}^{\infty} {b_{12,j}z_2(t-j)} + \varepsilon_1 (t), \\
     z_2(t) & = \sum_{j=1}^{\infty} {b_{21,j}z_1(t-j)}+\sum_{j=1}^{\infty} {b_{22,j}z_2(t-j)} + \varepsilon_2 (t)
\end{split}
\end{equation}

Rewriting the above equation in terms of the lag operator,

\begin{equation}
    \begin{bmatrix}
    b_{11}(L) & b_{12}(L) \\
    b_{21}(L) & b_{22}(L)
    \end{bmatrix}
    \begin{bmatrix}
    z_1(t) \\
    z_2(t)
    \end{bmatrix} =
    \begin{bmatrix}
    \varepsilon_1 \\
    \varepsilon_2
    \end{bmatrix},
    \label{eq_coupled_AR_lag}
\end{equation}
where,
$b_{ij}(L)= \sum_{k=0}^{\infty} b_{ij,k}L^k$, with $b_{ij,0}=\delta_{ij}$ or the Kronecker delta function. The covariance matrix of noise terms is:
\begin{equation}
    \boldsymbol{\Sigma}=
    \begin{bmatrix}
    \Sigma_{11} & \Sigma_{12} \\
    \Sigma_{21} & \Sigma_{22}
    \end{bmatrix},
\end{equation}
where, $\Sigma_{11}=\text{var}(\varepsilon_1)$, $\Sigma_{22}=\text{var}(\varepsilon_2)$ and $\Sigma_{12}=\Sigma_{21}=\text{cov}(\varepsilon_1, \varepsilon_2)$.

Taking Fourier transform of Eq. (\ref{eq_coupled_AR_lag}) on both sides:
\begin{equation}
    \begin{bmatrix}
    B_{11}(f) & B_{12}(f) \\
    B_{21}(f) & B_{22}(f)
    \end{bmatrix}
    \begin{bmatrix}
    Z_1(f) \\
    Z_2(f)
    \end{bmatrix} =
    \begin{bmatrix}
    E_1(f) \\
    E_2(f)
    \end{bmatrix}.
    \label{eq_coupled_AR_freq}
\end{equation}

In terms of transfer function matrix $\mathbf{H}(f)=[B_{ij}(f)]^{-1}$, Eq.~(\ref{eq_coupled_AR_freq}) can be expressed as:
\begin{equation}
        \begin{bmatrix}
    Z_1(f) \\
    Z_2(f)
    \end{bmatrix} =
    \begin{bmatrix}
    H_{11}(f) & H_{12}(f) \\
    H_{21}(f) & H_{22}(f)
    \end{bmatrix}
    \begin{bmatrix}
    E_1(f) \\
    E_2(f)
    \end{bmatrix}.
    \label{eq_coupled_AR_transfer}
\end{equation}

The spectral density matrix for the above is given by:
\begin{equation}
    \mathbf{S}(f)=\mathbf{H}(f)\boldsymbol{\Sigma} \mathbf{H}^*(f),
\end{equation}
where $*$ denotes matrix adjoint. To check the causal influence from $z_2$ to $z_1$ using spectral Granger Causality, we need to look at the autospectrum of $z_1$, which is given by:
\begin{equation}
\begin{split}
    S_{11}(f)=H_{11}(f)\Sigma_{11} H_{11}^*(f) + 2\Sigma_{12}\text{Re}(H_{11}H_{12}^*) \\ + H_{12}(f)\Sigma_{22} H_{12}^*(f).
\end{split}
\end{equation}

In the above expression, due to the cross terms, causal power contribution to $z_1$ is not explicit. Geweke, in his work~\cite{geweke1982measurement}, came up with a transformation that makes the causal power term and the intrinsic term explicit. To obtain this transformation for $z_1$, Eq.~(\ref{eq_coupled_AR_freq}) is left multiplied on both sides by: 
\begin{equation*}
\begin{bmatrix}
1 & 0 \\
-\Sigma_{12}/\Sigma_{11} & 1
\end{bmatrix}.
\end{equation*}

This gives,
\begin{equation}
    \begin{bmatrix}
    B_{11}(f) & B_{12}(f) \\
    \tilde{B}_{21}(f) & \tilde{B}_{22}(f)
    \end{bmatrix}
    \begin{bmatrix}
    Z_1(f) \\
    Z_2(f)
    \end{bmatrix} =
    \begin{bmatrix}
    E_1(f) \\
    \tilde{E}_2(f)
    \end{bmatrix},
\end{equation}

where, $\tilde{B}_{21}(f)=B_{21}(f)-\frac{\Sigma_{12}}{\Sigma_{11}}B_{11}(f)$, $\tilde{B}_{22}(f)=B_{22}(f)-\frac{\Sigma_{12}}{\Sigma_{11}}B_{12}(f)$. Let the new transfer function be given by $\mathbf{\tilde{H}}(f)$, whose elements then become $\tilde{H}_{11}(f)=H_{11}(f)+\frac{\Sigma_{12}}{\Sigma_{11}}H_{12}(f)$, $\tilde{H}_{12}(f)={H}_{12}(f)$, $\tilde{H}_{21}(f)=H_{21}(f)+\frac{\Sigma_{12}}{\Sigma_{11}}H_{22}(f)$ and $\tilde{H}_{22}(f)={H}_{22}(f)$. $\tilde{E}_{2}(f)={E}_{2}(f)-\frac{\Sigma_{12}}{\Sigma_{11}}{E}_{1}(f)$ which results in the $cov({E}_{1},\tilde{E}_{2})=0$ and $\tilde{\Sigma}_{22}$, the new variance of the error term of $z_2$, is $\Sigma_{22}-\frac{\Sigma_{12}^2}{\Sigma_{11}}$. 

As $\tilde{\Sigma}_{12}$ becomes zero, the autospectrum of $z_1(t)$ is now decomposed into two explicit parts:
\begin{equation}
    \tilde{S}_{11}(f)=\tilde{H}_{11}(f)\Sigma_{11} \tilde{H}_{11}^*(f) + H_{12}(f)\tilde{\Sigma}_{22} H_{12}^*(f).
\end{equation}

The first term in the decomposition of $\tilde{S}_{11}(f)$ represents an intrinsic power term of $z_1(t)$ and the second is due to the `causal power' contribution from $z_2(t)$. Spectral Granger Causality at a particular frequency $f$ is given by the logarithm of the ratio of total power to intrinsic power, so GC from $z_2$ to $z_1$ at $f$:
\begin{equation}
\label{eq_spectral_GC}
    I_{z_2 \rightarrow z_1}(f)=\ln{\frac{\tilde{S}_{11}(f)}{\tilde{S}_{11}(f)-(\Sigma_{22}-\frac{\Sigma_{12}^2}{\Sigma_{11}})|H_{12}(f)|^2}}.
\end{equation}

Multiplying Eq.~(\ref{eq_coupled_AR_freq}) with a constant, $G_f$, on both sides:
\begin{equation}
    \begin{bmatrix}
    B_{11}(f) & B_{12}(f) \\
    B_{21}(f) & B_{22}(f)
    \end{bmatrix}
    \begin{bmatrix}
    G_f Z_1(f) \\
    G_f Z_2(f)
    \end{bmatrix} =
    \begin{bmatrix}
    G_f E_1(f) \\
    G_f E_2(f)
    \end{bmatrix}.
\end{equation}

This gives (using Eq.~(\ref{eq_scaled_freq_comps})),
\begin{equation}
    \begin{bmatrix}
    B_{11}(f) & B_{12}(f) \\
    B_{21}(f) & B_{22}(f)
    \end{bmatrix}
    \begin{bmatrix}
    Y'_1(f) \\
    Y'_2(f)
    \end{bmatrix} =
    \begin{bmatrix}
    G_f E_1(f) \\
    G_f E_2(f)
    \end{bmatrix}.
\end{equation}

The above gives us a relation between the frequency components of $y'_1$ and $y'_2$ which can be thought of as being compressed counterparts of $z_1$ and $z_2$. If $z_1$ and $z_2$ are compressed by the same compressed sensing matrix, their coefficients corresponding to a particular frequency will be scaled by the same constant. This follows directly from Eq.~(\ref{eq_scaled_freq_comps}).

Hence, the transformation of $z_1$ to $y'_1$ and of $z_2$ to $y'_2$, changes only the variance of $z_1$, $z_2$, basically scaling $E_1(f)$ and $E_2(f)$ by the same quantity. Thus the variance and covariance terms of $E_1$ and $E_2$ are all scaled by $G_f^2$. To compute the causal influence from $y'_2$ to $y'_1$, the same procedure holds as for $z_2$ to $z_1$, with the only difference that both the numerator and the denominator in the logarithmic term of Eq.~(\ref{eq_spectral_GC}) are multiplied by $G_f^2$ as all the error variance and covariance terms are scaled by this quantity. As this scaling factor exists both in the numerator and denominator, it gets cancelled out, yielding the same causal influence from $Y'_2(f) \rightarrow Y'_1(f)$ as exists between $Z_2(f) \rightarrow Z_1(f)$. Thus,

\begin{equation}
    I_{y'_2 \rightarrow y'_1}(f)= I_{z_2 \rightarrow z_1}(f)
\end{equation}

Geweke~\cite{geweke1982measurement} showed that for all processes of practical interest, 
\begin{equation}
    \frac{1}{2\pi}\int_{-\pi}^{\pi}I_{z_2 \rightarrow z_1}(f) \,df= F_{z_2 \rightarrow z_1},
\end{equation}
where, $F_{z_2 \rightarrow z_1}$ refers to the time-domain GC from $z_2$ to $z_1$. Now, since $z_1$ and $z_2$ are sparse and all information contained in them is equally well represented in the compressed sensing domain by $y_1$ and $y_2$ (invertability/ reconstruction of all sparse vectors is guaranteed by the RIP property), all the information including causal information is present in the limited/ coarser frequency components of the compressed signal. Hence,
\begin{equation}
    \frac{1}{2\pi}\int_{-\pi}^{\pi}I_{z_2 \rightarrow z_1}(f) \,df= \frac{1}{2\pi}\int_{-\pi}^{\pi}I_{y'_2 \rightarrow y'_1}(f) \,df = F_{y'_2 \rightarrow y'_1},
\end{equation}
showing that the time-domain GC is preserved.

 To find the causal influence from $z_1$ to $z_2$ or from $y'_1$ to $y'_2$ at any frequency $f$, one can multiply by the transformation matrix,
$\begin{bmatrix}
1 & -\Sigma_{12}/\Sigma_{22} \\
0 & 1
\end{bmatrix}$, to delineate the intrinsic and the causal terms.

\subsection*{Toeplitz Sensing Matrices}

When our compressed signal is obtained by multiplication with a toeplitz sensing matrix, the following discussion holds. Let us start by considering an $n \times n$ toeplitz matrix, $T_n$.

\begin{equation}
T = 
\begin{bmatrix}
a_{0} & a_{n-1} & a_{n-2} & \cdots & a_{1} \\
a_{-1} & a_{0} & a_{n-1} & \cdots & a_{2} \\
a_{-2} & a_{-1} & a_{0} & \cdots & a_{3} \\
\vdots  & \vdots & \vdots & \ddots & \vdots  \\
a_{-(n-1)} & a_{-(n-2)} & a_{-(n-3)} & \cdots & a_{0}
\end{bmatrix}_{n \times n}.
\end{equation}
where, $a_i \sim \mathcal{N}(0,1)$. Multiplying $T$ with a signal $z$ gives,
\begin{equation}
    y_{n \times 1}= T_{n \times n}z_{n \times 1}
    \label{eq_CS_toep}
\end{equation}

This multiplication can be achieved by embedding $T_n$ into a circulant matrix, $C_{2n}$, of dimension $2n \times 2n$. This embedding is done in such a manner that the first $n$ entries of $C_{2n}z$ will be equal to $T_{n}z$:

\begin{equation}
\label{eq_toep_circ_form}
    \begin{bmatrix}   
    y \\
    y_d
    \end{bmatrix} = C_{2n} \begin{bmatrix}
    z \\
    0
    \end{bmatrix} 
     =\begin{bmatrix}
    T_n & B_n \\
    B_n & T_n
    \end{bmatrix}
    \begin{bmatrix}
    z \\
    0
    \end{bmatrix}
   = \begin{bmatrix}
    T_{n}z \\
    B_{n}z
    \end{bmatrix}.
\end{equation}

Here, the signal $z$ is appended with $n$ zeroes, resulting in a $2n \times 1$ vector. Once again, this circulant matrix is diagonalizable by a DFT matrix. Then, as in Eqs.~(\ref{eq_circ_fourier1}) and~(\ref{eq_full_circ_freqs}),

\begin{equation}
\label{eq_toep_freqs}
\begin{split}
       \begin{bmatrix}   
    y \\
    y_d
    \end{bmatrix} &= F_{2n}^{-1}DF_{2n} \begin{bmatrix}
    z \\
    0
    \end{bmatrix} \\
    & = F_{2n}^{-1} \begin{bmatrix}
    G_0Z(f_0) \\
    G_1Z(f_1) \\
    G_2Z(f_2) \\
    \vdots \\
    G_{2n-1}Z(f_{2n-1})
    \end{bmatrix},
\end{split}
\end{equation}
where, $Z(f)$ is the fourier coefficient at frequency $f$ of $\begin{bmatrix}
    z \\
    0
    \end{bmatrix}$. The $G$s here are the coefficients of the DFT of the first column in the entries of $C_{2n}$, given by, $\vec{a}_{2n}=[a_0, a_{-1}, \ldots, a_{-(n-1)}, a_0, a_{1}, \ldots, a_{(n-1)}]$.
    
It is known that zero padding a (finite length) signal only increases its frequency resolution without loosing any information in the frequency content of the original signal $z$. 

Now, we multiply Eq.~(\ref{eq_toep_circ_form}) on both sides with projection matrix $P$ of order $m \times 2n$ (with $m<n$),
\begin{equation}
    P_{m \times 2n}\begin{bmatrix}   
    y \\
    y_d
    \end{bmatrix} =
    P_{m \times 2n} \begin{bmatrix}
    T_{n}z \\
    B_{n}z
    \end{bmatrix},
\end{equation}
which can alternatively be written as:
\begin{equation}
    P_{m \times n} y = y'=
    P_{m \times n} T_{n}z =T_{m \times n}z,
\end{equation}
as $m<n$. The above equation can be seen as left multiplying $z$ with a compressed sensing matrix $T_{m \times n}$ with toeplitz structure. Going back to the representation of $y$ in Eq.~(\ref{eq_toep_freqs}),
\begin{equation}
\begin{split}
     P_{m \times 2n}\begin{bmatrix}   
    y \\
    y_d
    \end{bmatrix} = P_{m \times n}y
    = P_{m \times 2n} F_{2n}^{-1} \begin{bmatrix}
    G_0Z(f_0) \\
    G_1Z(f_1) \\
    G_2Z(f_2) \\
    \vdots \\
    G_{2n-1}Z(f_{2n-1})
    \end{bmatrix}.
\end{split}
\end{equation}

Multiplying on both sides by $F_m$,
\begin{equation}
    F_my'
    = P_{m \times 2n} \begin{bmatrix}
    G_0Z(f_0) \\
    G_1Z(f_1) \\
    G_2Z(f_2) \\
    \vdots \\
    G_{2n-1}Z(f_{2n-1})
    \end{bmatrix},
\end{equation}
results in,
\begin{equation}
    \begin{bmatrix}
    Y'(f_0) \\
    Y'(f_1) \\
    \vdots \\
    Y'(f_{m-1})
    \end{bmatrix} =
    \begin{bmatrix}
    G_0Z(f_0) \\
    G_1Z(f_1) \\
    \vdots \\
    G_{m-1}Z(f_{m-1})
    \end{bmatrix}.
\end{equation}

As the frequency components of $y'$ are scaled versions of the frequency components of $z'$, as seen for circulant compressed sensing matrices, the entire argument of GC preservation in the compressed sensing domain follows here as well. 

Thus, we have proved that GC is preserved if sensing matrices have circulant or toeplitz structure.


\section{Results}
\label{section_res}
\subsection{Simulations}

Sparse input signals, $z_1$ and $z_2$, were first simulated. These were then passed as an input to the same sensing matrix $C$ in order to obtain compressed signals, $y_1$ and $y_2$ respectively. The length of input signals was kept as $n=2000$ and that of the compressed signals was kept as $m=200$. Sparsity of the input signals, $k$, was kept as 20, except in the cases where results were specifically obtained for varying $k$.

Let the signals $z_1$ and $z_2$ be defined over a set $T=\{1,2,3,\ldots,N\}$, which can be considered to be a set of time points. Let $T_1 \subset T$ such that $T_1$ consists of $k$ elements chosen from $T$ in a uniformly random manner. Also, let $T_2 \subset T$ be defined as $T_2=\{t_2:t_2=t_1+1, \forall \: t_1 \in T_1\}$. $T_1^c$ and $T_2^c$ denote the complement sets of $T_1$ and $T_2$ respectively ($T$ being the universal set). 
The signals $z_1$ and $z_2$ were generated by \emph{sparsifying}  a pair of autoregressive processes $Z_1$ and $Z_2$. These were generated as per the following equations:
\begin{equation}
\begin{aligned}
  Z_1 (t)= & \alpha Z_1 (t-1) + \varepsilon_1 (t), \quad \forall t \in T, \\	       
  z_1 (t) = & \begin{cases} Z_1 (t), \quad \forall \: \{t \in T_1\}, \\
  0, \qquad \forall \: \{t \in T_1^c\},
  \end{cases}
 \end{aligned}
  \label{eq_sparse_AR_ind}
\end{equation}

\begin{equation}
\begin{aligned}
Z_2 (t)= &\beta Z_2 (t-1)+ \gamma z_1 (t-1)+\varepsilon_2 (t), \quad \forall t \in T, \\
z_2 (t)= & \begin{cases}
Z_2(t), \quad \forall \{t \in T_2\}, \\
0, \quad \forall \{t \in T_2^c\},
\end{cases}
\end{aligned}
\label{eq_sparse_AR_dep}
\end{equation}

\noindent where $\varepsilon_1$ and $\varepsilon_2$ are independent Gaussian noise drawn from $\mathcal{N}(0,0.1)$, $\alpha=0.8$, $\beta=0.08$ and $\gamma=0.75$. At $k$ randomly selected time points, $z_1(t)$ takes the values of $Z_1(t)$ and is set to zero elsewhere (at the rest $N-k$ time points). $z_2(t)$ is zero whenever $z_1 (t-1)=0$. At the rest $k$ time points, $z_2(t)$ retains the values of $Z_2(t)$. Defined this way, both $z_1$ and $z_2$ are $k$-sparse signals. As can be seen from the above equations, $z_1$ has a causal influence on $z_2$ (with coupling coefficient $\gamma$, controlling the strength of causation), and there is no causal influence from $z_2$ to $z_1$. 
%

\subsubsection{Performance of Structured Matrices}

\begin{figure}[b]
    \centering
    \includegraphics[width=0.98\columnwidth, trim={0 0 0 .01cm},clip]{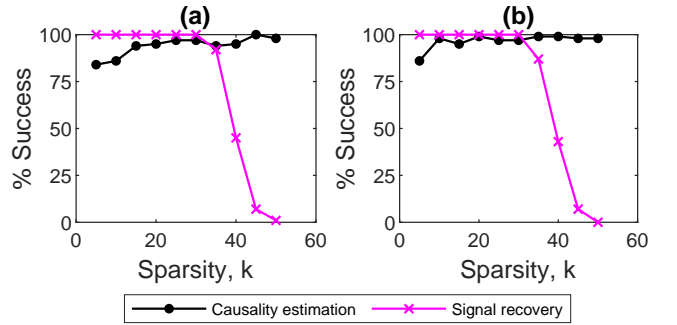}
    \caption{Percentage success (using 100 realizations) in sparse signal reconstruction and causality estimation for varying sparsity ($k$) in case of (a) circulant and (b) toeplitz sensing matrices. Signal recovery degrades for higher values of $k$, however, causality estimation improves.}
    \label{fig_toep_circ_varying_sparsity}
\end{figure}

Performance of circulant and toeplitz matrices was evaluated for sparse signal recovery and causality estimation, while varying the level of sparsity, $k$. 100 realizations of $z_1$ and $z_2$ with random initial values were simulated. Percentage success in signal reconstruction and correct causality detection is shown in Fig.~\ref{fig_toep_circ_varying_sparsity}. $k$ was varied from 5 to 50 in steps of 5. CVX, a MATLAB package for convex optimization~\cite{grant2014cvx} was used for $l_1$ minimization and subsequent sparse signal reconstruction for all results in this paper. For each realization, sparse signal reconstruction ($z_1$ and $z_2$ from $y_1$ and $y_2$ respectively) is counted as successful if the mean squared error between the original and reconstructed signal is less than $10^{-5}$ for both $z_1$ and $z_2$. The mean squared error of reconstruction for a realization of $z_1$ is estimated as:
\begin{equation}
    MSE(z_1)= \frac{1}{N}\sum_{i=1}^{N}(z_1(i)-\hat{z}_1(i))^2,
    \label{eq_MSE}
\end{equation}
where, $i$ denotes the temporal index of samples in $z_1$, $N$ is the length of $z_1$ and $\hat{z}_1$ is the reconstructed signal for $z_1$. For each realization causality estimation is counted as being successful if the Granger causality F-statistic from $y_1$ to $y_2$ is found to be significant and from $y_2$ to $y_1$ is found to be insignificant. Multivariate Granger Causality (MVGC) toolbox~\cite{mvgc} was used for the computation of GC F-statistic and its significance for all experiments in this work. The order of the AR processes was determined using Akaike Information Criterion with maximum number of lags to be considered for the processes set to 30. Significance testing was done using $\chi^2$ method with significance level, $\alpha$, set to $0.01$. Rest of the parameters of the toolbox were set to their default values~\cite{mvgc}. These settings for the computation of GC and its significance were kept constant for all experiments in this paper.

\begin{figure}[h]
    \centering
    \includegraphics[width=0.6\columnwidth, trim={0 0 0 .1cm},clip]{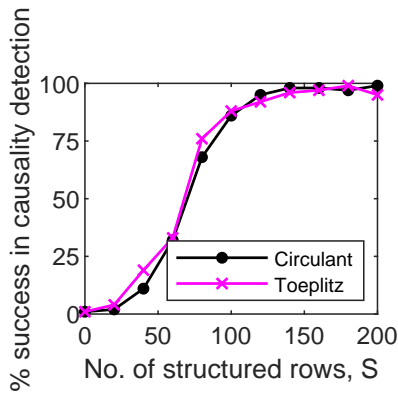}
   \caption{Percentage success (using 100 realizations) in correct causality estimation for varying number of structured rows, $S$, in case of circulant and toeplitz sensing matrices. Causality estimation improves as the number of structured rows increases for both types of matrices.}
    \label{fig_toep_circ_varying_struc_rows}
\end{figure}

Percentage success in causality estimation when the level of sparsity is kept constant at $k = 20$ but the degree of circulant and toeplitz structure is varied, is shown in Fig.~\ref{fig_toep_circ_varying_struc_rows}. The degree of structure was varied by allowing only the first $S \leq m$ rows of the sensing matrix to have circulant/ toeplitz structure while the entries for the rest of $m-S$ rows were selected randomly from $\mathcal{N}(0,1)$. $S$ was varied from 0 to $m (=200)$ in steps of 20. For each $S$, 100 realizations of $z_1$ and $z_2$ were simulated and compressed to obtain $y_1$ and $y_2$ respectively (using a different sensing matrix for each realization). Reconstruction of $z_1$ and $z_2$ was found to be successful for all the realizations for all the values of $S$ here. Hence, the depiction of percentage success in sparse signal recovery is omitted in Fig.~\ref{fig_toep_circ_varying_struc_rows}. 

\subsubsection{Varying the Coefficient of Causation}

\begin{figure}
    \centering
    \includegraphics[width=0.95\columnwidth]{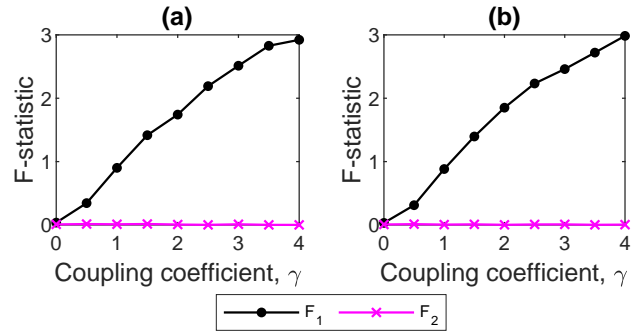}
    \caption{Mean Granger Causality values (using 100 realizations) from $y_1 \rightarrow y_2$ ($F_1$) and $y_2 \rightarrow y_1$ ($F_2$) as the coupling coefficient $\gamma$ is varied in case of sensing by (a)~circulant and (b) toeplitz matrices. With increasing $\gamma$, strength of GC estimated in the direction of coupling between the compressed signals increases.}
    \label{fig_toep_circ_varying_coeff}
\end{figure}

For simulated input signals $z_1$ and $z_2$, the coupling coefficient $\gamma$ (see Eqs.~(\ref{eq_sparse_AR_ind}) and (\ref{eq_sparse_AR_dep})) was varied from 0 to 4 in steps of 0.5. GC F-statistic was estimated from $y_1$ to $y_2$ ($F_1$) and from $y_2$ to $y_1$ ($F_2$) for both fully circulant and toeplitz sensing matrices. In Figs.~\ref{fig_toep_circ_varying_coeff}(a) and \ref{fig_toep_circ_varying_coeff}(b), mean values of $F_1$ and $F_2$, obtained using 100 random realizations are shown for circulant and toeplitz matrices respectively as the coefficient $\gamma$ is varied.

\subsubsection{Multivariate neural spike trains}

\begin{figure*}
    \centering
    \includegraphics{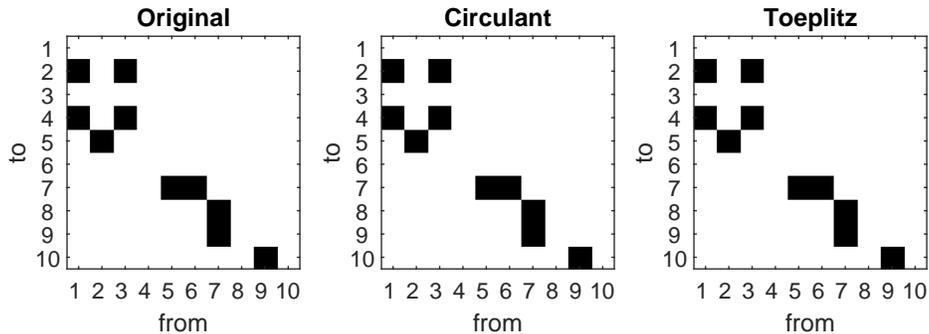}
    \caption{Granger Causality connectivity matrices recovered using circulant and toeplitz sensing matrices for a simple simulated network of 10 spiking neurons using 20 realizations of the processes. The leftmost subfigure represents the original connectivity matrix. The original connections/ significant connections are represented with black squares and no connections are represented in white.}
    \label{fig_simple_simu_res}
\end{figure*}

A simple network of simulated spike trains from 10 neurons was generated as in~\cite{zhang2016statistical}. MATLAB scripts from~\cite{zhang2016statistical} are openly available and were directly used for the purpose. These spike trains are based on the Generalized Linear Model (GLM). The connectivity matrix for this network is as shown in Fig.~\ref{fig_simple_simu_res} (leftmost plot). Excitatory and inhibitory connections are not differentiated in the displayed connectivity matrix. 20 realizations with each of the 10 nodes generating spike trains of length 10000 time points were simulated. 

Each of the spike trains were compressed using circulant and toeplitz sensing matrices with $m=2000$. $m$ was chosen to be greater than four times the sparsity of the signal with maximum sparsity. Conditional MVGC was estimated using the MVGC toolbox with the parameters of the toolbox set as discussed before. The connectivity matrix recovered from the compressed signal using circulant and toeplitz sensing matrices is shown in Fig.~\ref{fig_simple_simu_res}. It is found that both types of matrices perfectly recover all the existing causal connections in the simulated network and do not detect any false positives. Thus, the sensitivity as well as the specificity of the estimated connections is found to be 1.

\subsection{Real Data}
\label{sec_CS_real_data}
Sparse neuronal spike train data, recorded from the prefrontal cortex of adult male Sprague-Dawley rats, acquired for the study in~\cite{zhang2016statistical}, are available open source (CC BY 4.0 License). These recordings are made from putative single units in the prelimbic region of the prefrontal cortex of the rats, while they performed a T-maze based delayed-alternation task of working memory. To describe the task briefly, rats (already trained for a number of trials) had to navigate the runway of a T-maze and choose one of the two arms opposite to the one visited last time in order to obtain food rewards given by the experimenter. Since the prefrontal cortex plays an important role in cognitive and behavioral processes, the construction of functional connectivity based neuronal networks can give insight into its working.

We use neuronal spike data for `Experiment 6' from the available dataset from~\cite{zhang2016statistical} and check for GC estimates between the considered network of 63 neurons. We analyze the entire time series of length $20,114$ time points from these neurons. Thus, $n=20114$, and $m$ is taken to be $7000$ (set to be greater than $2$ times the sparsity of the signal having the maximum sparsity).

In~\cite{zhang2016statistical}, a `Structural Information Enhanced' regularization method has been developed in order to aid the GLM framework to better estimate functional connectivity between neurons. This technique is mainly for large sparse spike train datasets. As discussed in the introduction, GLM is a model-based method. Simulation results in~\cite{zhang2016statistical} indicate that when the parameter selection for GLM is done using the proposed regularized method (abbreviated as SGL), the technique performs better than existing approaches. As a result, the authors display a confidence in functional connectivity estimated in the rats' prefrontal cortex using the above discussed data recordings. However, even for the simple network simulation of $10$ neurons discussed in the previous section, the method does not perfectly estimate the network connections. For the simulation case considered in this manuscript, the best that the method does is to discover an average $7.23$ correct connections (that is, true positives) out of the existing 10 connections, using a $100$ realizations of the simulation. This result is given in Table 1 of~\cite{zhang2016statistical}. Our method, on the other hand, recovers the connectivity matrix perfectly (see Fig.~\ref{fig_simple_simu_res}). It is found that the SGL method has high specificity but low sensitivity. In other words, most truly zero-valued connections are estimated to be zero, however, the sensitivity of detecting significant connections is not as good.

\begin{figure*}
    \centering
    \includegraphics{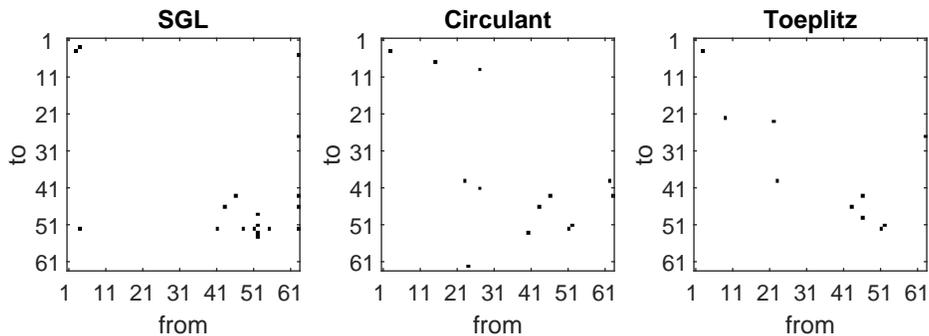}
    \caption{Granger Causality connectivity matrices recovered using circulant and toeplitz sensing matrices for a real network comprised of 63 networks, with spike trains recorded from a rat prefrontal cortex~\cite{zhang2016statistical}. The leftmost subfigure represents the connectivity matrix obtained using `SGL' method~\cite{zhang2016statistical}. The significant connections are represented with black squares and no connections are represented in white.}
    \label{fig_expt6_res}
\end{figure*}

For the real data used in this section, we compare the performance of fully circulant and toeplitz sensing matrices to discover causal connections for the network with the connectivity discovered by SGL method. This was done as there is no ground truth available to compare with. Fig.~\ref{fig_expt6_res} shows the connectivity matrices obtained by the three methods. Significant connections are displayed as black squares. Any non-zero couplings discovered by SGL are displayed as significant, while conditional GC estimation and its significance detection were done by the MVGC toolbox using the same parameters as discussed for the simulations before. The SGL results are basically a reproduction of the results for `Experiment 6', displayed in Fig. 10 of~\cite{zhang2016statistical}. True negative rate or specificity of the results using circulant and toeplitz sensing matrices with respect to the SGL results were found to be 0.998 and 0.999 respectively. True positive rate or sensitivity was found to be 0.333 for both circulant and toeplitz sensing. As discussed before, for simulations, SGL showed high specificity but relatively low sensitivity. Thus, we can expect the relative specificity of our method obtained here to be close to the true specificity. On the other hand, we would expect the true sensitivity to be much better than the relative sensitivity obtained here as our method showed a specificity and sensitivity of 1 for the simulation network, which is superior to the performance of SGL for the simulation.    

\section{Discussion and Conclusions}
\label{section_discussion}
This work discusses the task of discovering causal connections for signals in the compressed sensing domain. To the best of our knowledge, ours' is the pioneering study to solve the problem of causality estimation between compressed measurement signals without the need to reconstruct sparse signals. Causal inference is fundamental to understanding the interaction and underlying dynamics of processes in many disciplines of science. With the widespread use of compressed sensing for acquisition, transfer and storage of signals, it becomes essential that causal analyses be performed in the compressed domain. This will save the computational cost of reconstruction of sparse signals. Furthermore, it is known that for sparse temporal data such as sparse point processes, it is not directly possible to estimate `model-free' causality for time-domain data. For such data as well, causal analysis directly in compressed domain or after transforming sparse signals to compressed domain can be very useful.

We have mathematically proved that structured compressed sensing matrices, specifically circulant and toeplitz matrices preserve GC for sparse signals in the compressed domain. These signals need not necessarily be sparse in the temporal domain and may possess a sparse representation in another domain. However, the processes must be covariance stationary in the temporal domain for GC to be applicable. GC is the oldest and most widely used method for data-driven causality estimation and has been applied in a large number of studies across disciplines. Our proof opens up numerous more useful applications of GC.  

To verify the proof and demonstrate its significance in practical applications, the performance of the structured sensing matrices is tested for GC based causality detection using a number of simulation experiments. For two coupled sparse processes, when the amount of signal sparsity $k$ is varied and the structure of sensing matrix is kept as circulant or toeplitz, percentage success in causality estimation increases and reaches $100\%$ at high values of $k$ ($>$ 30). On the other hand, percentage success in signal reconstruction deteriorates with increasing values of $k$. This can be seen from Fig.~\ref{fig_toep_circ_varying_sparsity}. When the input signal sparsity level is constant, but the number of structured rows in the sensing matrix are increased, then as seen in Fig.~\ref{fig_toep_circ_varying_struc_rows}, the percentage success in causality estimation increases and approaches $100\%$ for fully structured (circulant or toeplitz) matrix. This illustrates that these structured sensing matrices are capable of preserving time domain GC owing to the preservation of spectral GC at different frequencies. As the number of structured rows in these matrices increase from zero to higher values, causal content at more and more frequencies is preserved, resulting in an improved causality detection performance. For a fully structured matrix, spectral GC at all frequencies is preserved.

Along with preserving the direction of causation, the structured sensing matrices used here also preserve the relative strength of causation in the compressed domain. This is evident from Fig.~\ref{fig_toep_circ_varying_coeff}, where the strength of estimated GC F-statistic in the actual causal direction increases in the compressed domain, when $\gamma$, the unidirectional coupling coefficient, is increased. In the direction in which there is no coupling, estimated causal strengths are observed to be close to zero in the compressed domain.

In the case of multivariate simulations of sparse neural spike trains, it is seen that the use of conditional GC on compressed signals helps to discover the connectivity structure of the network perfectly for both circulant and toeplitz structured sensing matrices. $100\%$ specificity and sensitivity of the technique is observed in this case and the performance of our method is shown to be superior to a GLM model-based method of discovering causal connectivity. Model-based methods are widely used for discovering functional (causal) connectivity from sparse neural spike trains, however may have limited scope because of specific model assumptions and high computational costs.

We also demonstrate a real world application of our method on neural spike train recordings from a rat prefrontal cortex, where the rat was performing a delayed-alternation task of working memory. These results shown in Fig.~\ref{fig_expt6_res} are promising, as many common connections are discovered by circulant and toeplitz sensing matrices and some of these overlap with the connections reported by a GLM based method, `SGL'~\cite{zhang2016statistical}. As our method shows better performance than the SGL on simulated multivariate data, the results obtained in this work can be considered to be more reliable. 

Networks that are required to be analyzed based on causal connections, and for which signals are acquired in the compressed domain, the use of structured sensing matrices would help to solve the causal estimation task within the compressed domain. Preservation of granger causality by circulant and toeplitz matrices would then be the deciding criteria for design of sensors in various systems. In fact, even currently, many measurement technologies impose structure on the matrix. This is because structured matrices possess other advantages, for example, requirement of lesser number of independent random variables for matrix generation and better efficiency of recovery algorithms as the matrix admits a fast matrix–vector multiply~\cite{bajwa2007toeplitz, rauhut2012restricted}. Causal inference in the compressed domain in these cases is a useful and powerful technique.

Also, in case of naturally occurring sensors that can be approximated to be sensing the signals based on operation by a structured matrix like partial or full circulant/toeplitz, causality analysis for the signals can be easily and reliably performed. One way by which the operation performed by a sensing matrix can be deciphered is by checking the properties of the sensed signal. For example, toeplitz matrices are known to perform a moving average operation on the input signals~\cite{gray2006toeplitz}. Some neural signals are known to be compressed signals~\cite{ganguli2012compressed, nagaraj2016neural} while most single unit neuronal signals are sparse spike trains~\cite{gerstner2002spiking}. Also, some neuroimaging modalities, such as the fMRI widely employ compressed sensing approaches for acquisition of signals~\cite{jung2007improved, zong2014compressed, tsao2012mri}. On a separate note, direct granger causal analysis of fMRI data has faced criticisms, one of the reasons for this being the unreliability of results due to low temporal resolution of acquired data~\cite{seth2015granger}. Hence, recognition and design of causality-preserving sensing matrices can prove to be extremely useful for analysing functional neural connections and brain connectivity.

It will be interesting to check if any other type of sensing matrices with binary/ real-valued entries preserve GC.  Further, it will be useful to test if other causality measures such as nonlinear variations of GC, Transfer Entropy~\cite{schreiber} and Compression Complexity Causality~\cite{kathpalia2019data}, preserve causality under circulant, toeplitz or other types of sensing matrices. We would also like to estimate GC directly from real compressively sensed signals acquired using structured sensing matrices discussed in this work and try to cross-verify the validity of the proposed scheme in such cases.

\begin{acknowledgments}
The authors would like to thank past interns at the National Institute of Advanced Studes: M.A. Abhijith, Aishwarya Nambissan, Aswathi Gopinath and M. Krishnapriya (from Amrita University) for running some preliminary simulations associated with this work. This study is supported by Czech Science Foundation, Project No.~GA19-16066S and by the Czech Academy of Sciences, Praemium Academiae awarded to M. Palu\v{s}. Nithin Nagaraj gratefully acknowledges the financial support of Cognitive Science Research Initiative, Dept. of Science \& Tech., Govt. of India (CSRI-DST) Grant No. DST/CSRI/2017/54(G).
\end{acknowledgments}


\bibliography{apssamp}

\end{document}